\documentclass{svproc}
\usepackage{url}

\usepackage{graphicx}

\usepackage{listings}
\usepackage[numbers]{natbib}

\lstdefinestyle{jsonstyle}{
    basicstyle=\small\ttfamily,
    breaklines=true,                 % Automatic line breaking
    keywordstyle=\color{blue},       % Style for keywords
    commentstyle=\color{green},      % Style for comments
    stringstyle=\color{red}          % Style for strings
}

\begin{document}
\mainmatter              % start of a contribution
%
%\title{Advanced computing for reproducibility of astronomy Big Data Science, with a showcase of AMIGA and the SKA Science prototype}
\title{Advanced computing for reproducibility of astronomy Big Data Science, with a showcase of AMIGA and the SKA Science prototype}

\titlerunning{Advance computing for reproducibility in SKA}  % abbreviated title (for running head)
%                                     also used for the TOC unless
%                                     \toctitle is used
%
\author{Julián Garrido\inst{1} \and Susana Sánchez\inst{1} \and Edgar Ribeiro João\inst{1} \and Roger Ianjamasimanana\inst{1} \and Manuel Parra\inst{1} \and Lourdes Verdes-Montenegro\inst{1}}
\authorrunning{Julián Garrido et al.} % abbreviated author list (for running head)
%
%%%% list of authors for the TOC (use if author list has to be modified)
\tocauthor{Julián Garrido, Susana Sánchez, Edgar João, Roger Ianjamasimanana, Manuel Parra and Lourdes Verdes-Montenegro}
\institute{Instituto de Astrofísica de Andalucía, CSIC, Spain,\\
\email{jgarrido@iaa.csic.es}
}

\maketitle              % typeset the title of the contribution

\begin{abstract}
%The abstract should summarize the contents of the paper
%using at least 70 and at most 150 words. It will be set in 9-point
%font size and be inset 1.0 cm from the right and left margins.
%There will be two blank lines before and after the Abstract. \dots
% We would like to encourage you to list your keywords within
% the abstract section using the \keywords{...} command.
The Square Kilometre Array Observatory (SKAO) faces unprecedented technological challenges due to the vast scale and complexity of its data. This paper provides an overview of research by the AMIGA group to address these computing and reproducibility challenges. We present advancements in semantic data models, analysis services integrated into federated infrastructures, and the application to astronomy studies of techniques that enhance research transparency. By showcasing these astronomy work, we demonstrate that achieving reproducible science in the Big Data era is feasible. However, we conclude that for the SKAO to succeed, the development of the SKA Regional Centre Network (SRCNet) must explicitly incorporate these reproducibility requirements into its fundamental architectural design. Embedding these standards is crucial to enable the global community to conduct verifiable and sustainable research within a federated environment.
\end{abstract}

\section{Introduction}\label{sec:intro}

The Square Kilometre Array Observatory (SKAO) is an intergovernmental organisation established in 2021 to build and operate new generation radio telescopes. The SKAO will comprise two telescope sites in Australia and South Africa that will generate unprecedented volumes of data, both in scale and complexity. For this reason, participating countries have agreed to create a distributed network of \textit{SKA Regional Centres} (SRCNet) to host the scientific archive, provide storage and computational resources, and offer scientific user support. These centres will act as the interface between the global user community and the observatory’s data ecosystem.

The rate and scale of SKA data transmission are estimated on 7.4~PB/s from the Central Signal Processor (CSP) to the Science Data Processor (SDP) in Australia, while 8.9~Tb/s will be transferred in South Africa. The Observatory will deliver an estimated 700~PB per year from the SDP to the SRCNet. As a consequence, it will not be feasible to preserve the raw telescope data. 

The two foundational documents required for approval of SKA Phase~1 construction explicitly identify Open Science and reproducibility as core principles. The SKA Phase~1 Construction Proposal states that “Open Science, based on the precept of making scientific research collaborative, transparent and accessible to all, is rooted in SKA’s foundational principles”~\cite{SKA_ConstructionProporsal}. The \textit{SKA Observatory Establishment and Delivery Plan} further highlights that reproducibility will be a key metric of the observatory’s scientific success~\cite{SKA_DeliveryPlan}. In both cases, particular emphasis is placed on the preservation of workflows and provenance information, in alignment with the FAIR principles (Findable, Accessible, Interoperable, and Reusable).

More recent documents emerging from the SRCNet initiative reinforce this commitment. The document describing high level requirements specifies that the SRCs must support Open Access and Provenance preservation. For example, the SRCs will enable users to provide public links to SKA Science Data Products. In addition, it also establishes that SRCs will be able to save the complete workflow and provenance associated with any Advanced Data Products, alowing that they are queried, viewed, and re-used~\cite{bolton2019ska}. Likewise, the project explicitly states that “The SRC Network will embrace FAIR and Open Science principles whenever possible and appropriate” \cite{bolton_srcnet2023}.

This unprecedented data volume raises fundamental challenges in diverse domains such as Big Data management, e-Science, remote visualisation, among others. 
Some of these challenges connect to the execution of scientific workflows. For instance, computation must be brought to where data resides to minimise unnecessary data movement. The SRCNet heterogeneity in terms of storage and computing solutions at each node requires interoperable strategies. To handle this, data lake technologies are being deployed, including a prototype based on \textit{Rucio} \cite{Barisits2019}, designed to manage heterogeneous storage infrastructures. Computational heterogeneity introduces additional complexity: different nodes will feature different capacities, architectures and analytical tools. Achieving reproducibility in such an environment requires portable workflows and a common software stack that can be instantiated across nodes with containerised environments and virtualisation technologies.

Smart resource management is therefore essential. Workflows may have specific computational, memory, or data-access requirements that need to be scheduled efficiently according to both resource availability and network topology. Intelligent resource allocation and federated execution models are being explored as key ingredients to support large-scale, data-intensive analyses. The International Virtual Observatory Alliance (IVOA) community is working towards an \textit{Execution Broker} specification that describes executable entities and their computational requirements, allowing clients to discover where and when a given workflow can be executed across distributed infrastructures. Complementarily, The SRCNet initiative is actively investigating this concept as part of its \textit{Execution Planner} approach to federated workflow execution.

Astronomy has a long-standing tradition of Open Access through the efforts of the IVOA, which has defined many of the standards enabling interoperable access to astronomical data and services. 
In parallel, the broader Open Science and Artificial Intelligence (AI) communities are developing guidelines to ensure that advanced computing and data-driven methods remain FAIR and reproducible. A recent Delphi study proposed the “Top 10 best practices for FAIR implementation in Machine Learning (ML) and AI”~\cite{osmenaj_2025_14932518}, based on expert input collected across 15 countries. Among the practices for which consensus was reached are the following: 1) ML/AI models should be described with rich metadata; 2) 
clear instructions on how to deploy the model should be provided. These recommendations are highly relevant for SKA and SRCNet, where AI and ML will play a central role in data reduction, classification, and science analysis pipelines.

The European project AI4EOSC (Artificial Intelligence for the European Open Science Cloud) is another example of this convergence between AI, Open Science and advanced infrastructures~\cite{Diaz2024253}. AI4EOSC delivers a platform and software stack (AI4OS) providing tools and services to support the full machine learning lifecycle, from model creation and training to deployment and monitoring in production on top of EOSC and compatible infrastructures. The platform emphasises openness and portability (fully open-source and designed to avoid vendor lock-in). 

At the same time, the rapid rise of generative AI introduces both opportunities and risks for Open Science. An exploratory analysis of the positive and negative impacts of generative AI across the spectrum of Open Science practices was recently published \cite{Hosseini2025_GenAI_OS}. Using the taxonomy of the UNESCO Recommendation on Open Science, they examine effects on Open Scientific Knowledge (including open access publications, open research data, open source software and source code, and open evaluation), Open Science infrastructures, Open engagement of societal actors, and open dialogue with other knowledge systems. Their analysis highlights that generative AI can both amplify openness and exacerbate concerns around integrity, transparency and equity, underscoring the importance of robust provenance, licensing, and reproducibility mechanisms. 

We have long been aware of the need to address the challenge of handling SKA data in order to extract scientific knowledge, and our commitment is that this is done not only efficiently, but also in a reproducible way. Consequently, we are particularly engaged in ensuring that Big Data Science carried out in the SRCs follows collaborative and Open Science practices. 

Over the past years, the AMIGA group (\url{https://amiga.iaa.csic.es}) at IAA-CSIC has conducted extensive research and development aimed at addressing reproducibility and Open Science challenges in preparation for SKA operations. As part of this effort, IAA-CSIC has started prototyping an SRC that will be fully aligned with Open Science principles, integrating FAIR data management, workflow preservation and provenance capture into its design. The present paper reviews a selection of initiatives undertaken by AMIGA in this context. Section~\ref{sec:datamodels} discusses data models to improve interoperability in the SKA Science platforms, analysed from both the SDP and SRCNet perspectives. Section~\ref{sec:services} explores advanced data analysis services using two complementary approaches: Software as a Service (SaaS) and Function as a Service (FaaS) on distributed computing infrastructures.  Section~\ref{sec:reprod} focuses on reproducibility in AMIGA science studies, highlighting two case studies demonstrating different strategies to achieve reproducibility of scientific results using existing tools. Finally, section~\ref{sec:conclusions} explains the conclusions of the paper. 

\section{Data models to improve interoperability in the SKA Science platforms}\label{sec:datamodels}

\subsection{Science Data Processor data model}\label{sect:prov-data-model}

Provenance data is a vital aspect of scientific study, required to ensure reproducibility, verify quality, reliability, traceability, and aid in the identification and location of errors within complex workflows \cite{GMV_REP_SDP_MTD20}. %GMV_MDL_REF19, 
Given the enormous volume of information processed by the SDP, a standardised method for generating provenance is crucial, particularly for data manipulation and complex scientific workflows where intermediate data products may often be discarded.

The aim of developing a specific provenance Data Model (DM) is to create a consistent standard that fulfils the needs of the SKA-SDP, supporting Open Science requirements where metadata must be sufficiently complete to allow external researchers to understand and reproduce the experiment with the same results.

The reference Provenance Data Model is the PROV-DM developed by the World Wide Web Consortium (W3C), 
but astronomy community has adapted this standard and it created the IVOA Provenance Data Model (DM) \cite{IVOADM19}. The IVOA Provenance DM include traceability of products (e.g., images, catalogues, documents), acknowledgement and contact information, quality and reliability assessment, error location identification, structured provenance metadata search, and reproducibility. The IVOA model is a specialisation of PROV-DM, focusing on key classes like Entities, Activities, and Agents to trace the sequence of events involved in a workflow. This model proved robust, having been tested in projects like the Cherenkov Telescope Array (CTA) pipeline (CTApipe) to register workflow stages. Nontheless, the IVOA adaptation is designed for tabular implementations, despite of being the W3C standard graph-based.

Among the different provenance data models that exist in the literature, it is worth highlighting the ProvONE Data Model, which was also developed as an extension of W3C PROV-DM specifically to handle computationally-intensive, scientific workflow-based experiments. ProvONE addresses three main aspects of provenance: prospective provenance (capturing the 'recipe' or workflow specification), retrospective provenance (recording already executed steps by software agents), and process provenance (recording the evolution of the program). ProvONE is highly effective in defining the fundamental information required to understand and analyse scientific workflow execution, specifically recording the low-level programs, ports, controllers, and channels used by each execution.

Our study concluded that while the IVOA Provenance DM is a robust, well-consolidated model suited to astronomy, it lacked explicit and clear representation for the fine-grained, low-level programs and executions necessary for the complex processing expected from the SKA-SDP \cite{GMV_MDL_SDP20} \cite{GMV_REP_SDP_MTD20}. Consequently, we proposed the SKA-ProvSDP Data Model as a variation of the IVOA Provenance DM, expanded by incorporating the workflow representation section extracted from the ProvONE Data Model \cite{GMV_MDL_SDP20}.

The SKA-ProvSDP model leverages the key components of the IVOA framework to establish traceability and accountability, and the Data Model is structured around five core classes.  Entities and Collections represent data objects such as products, inputs, tables, or data models. Collections are used to group constituent Entities. Activities describe interactions with these Entities, such as processing or data transformation, and establish the input (\texttt{Used}) and output (\texttt{WasGeneratedBy}) relationships. Agents define responsibility for an Entity or Activity (e.g., a person or institution), and are crucial for citation and verification, ideally incorporating Persistent Identifiers (PIDs). Finally, Description Classes facilitate the reuse of common information (like file formats), and the Activity Configuration registers the parameters and files used to execute a process, which is vital for assessing reproducibility.

The provenance model is expanded by introducing classes such as Program, Port, Channel, and Controller to accurately represent complex workflow mechanics and low-level processing steps. The Program class represents the tasks performed, which can be broken down further using the \texttt{hadSubProgram} relation. Port and Channel manage the movement of Entities between different programs. Crucially, the SKA-ProvSDP Data Model associates an Activity with the specific Program that executed it via the \texttt{Executed} relation. This segregation of high-level Activity classes from low-level Program classes resolves compatibility conflicts inherent in simpler models, thus enhancing clarity and facilitating the representation of complex, granular processing structures similar to the Actor Model.

When implementing the core model, best practices suggest using Persistent Identifiers (PIDs) like Digital Object Identifiers (DOI) and Open Researcher and Contributor IDs (ORCID) to provide actionable, long-lasting references for entities and agents, resolving potential inconsistencies in naming or location over time. %Furthermore, efficient implementation of the model involves defining \texttt{ActivityDescription} files for shared processes that can be reused across multiple activities, reducing redundancy \cite{GMV_REC_SDP20}.

In summary, the SKA-ProvSDP model offers a robust, flexible, and comprehensive method for recording provenance in the SKA-SDP. It successfully integrates the astronomical specialisation of the IVOA Provenance Data Model with the critical workflow granularity provided by ProvONE, ensuring that both high-level pipeline activities and low-level computational steps are accurately and clearly documented for future reproducibility and analysis \cite{GMV_MDL_SDP20}.

%%%%%%%%%%%%%%%%%%%%%%%%%%%%%%%%%%%%%%

\subsection{The SRCNet data model}
\label{sect:srcnet-data-model}

\begin{figure*}[htbp]
    \centering
    \includegraphics[width=0.95\textwidth]{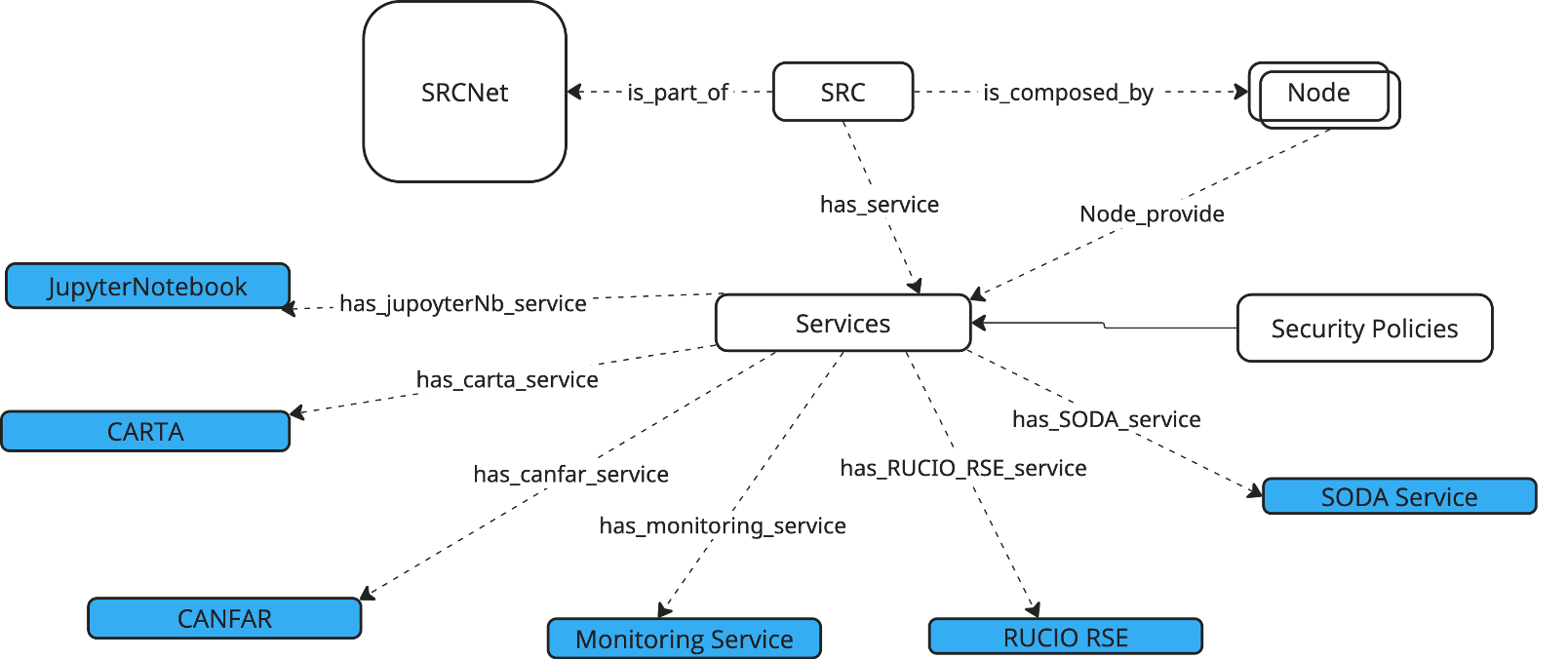}
    \caption{Conceptual view of the SRCNet data model showing the relationships between SRCNet, SRCs, Nodes, and Services. Services include Jupyter Notebook, CARTA, CANFAR, Rucio SE, Monitoring, and SODA.}
    \label{fig:srcnet-concepts}
\end{figure*}

A semantic model for the SRCNet is essential to describe its architecture, resources, and workflows in a machine-actionable manner. This model constitute an ongoing effort to support interoperability across SRC nodes and facilitates advanced functionalities such as AI-driven brokering and workflow orchestration \cite{EdgarSemanticModel2025}. Figure~\ref{fig:srcnet-concepts} illustrates a simplified view of the conceptual structure of the SRCNet, including the main entities (\textit{SRCNet}, \textit{SRC}, \textit{Node}, and \textit{Service}) and their relationships. Services include interactive analysis tools (e.g., Jupyter Notebook), visualisation platforms (e.g., CARTA), data management systems (e.g., Rucio Storage Element), and monitoring and data-access services such as SODA.

The SRCNet data model is structured around two complementary dimensions:

\begin{itemize}
    \item \textbf{Static Model:} Defines persistent entities such as SRCs, Nodes, and Services, along with their attributes and relationships. Each \textit{Node} encapsulates computing, storage, and networking resources, as well as local services.
    \item \textbf{Dynamic Model:} Captures time-dependent properties such as resource availability, service status, and performance metrics, enabling real-time orchestration and optimisation.
\end{itemize}

Nodes are described in terms of their hardware capabilities (CPU, GPU, RAM, disk), storage capacity, and network characteristics (bandwidth, latency). Services are linked to nodes through explicit relationships and include both global services (e.g., authentication, registry) and local services (e.g., notebooks, visualisation, monitoring).

The model is implemented using JSON-LD \cite{jsonld11} \cite{rfc2026}, a lightweight format for Linked Data that supports interoperability and integration with web technologies. Listing~\ref{lst:cpu-definition} shows an example of defining a EnergyLabel component within a node.

\begin{lstlisting}[style=jsonstyle, caption={Definition of energy label for an SRC node using JSON-LD}, label={lst:cpu-definition}]

 "has-energyLabel":{
            "@type":"",
            "has-energySource":{
              "@type":"",
              "renewableEnergyFactor":"",
              "energyReuseFactor":""
            },
            "has-energyEfficiency":{
              "@type":"",
              "energyEfficiencyPUE":"",
              "waterEfficiencyWUE":""
            }
\end{lstlisting}
            
This representation enables the semantic description of computing resources, which can be queried and reasoned upon using SPARQL in a semantic database such as Apache Jena Fuseki.

By formalising the SRCNet architecture and its operational characteristics, the semantic model lays the foundation for a future service broker capable of dynamic resource allocation and workflow scheduling. The broker will leverage both static and dynamic components of the model to optimise execution plans based on real-time availability and data locality.

Critical aspects for the SRCNet include the connectivity between the different nodes, the data distribution and locality. Replicas per dataset and local data dependencies (services requiring data to be processed at its storage location when transfer is infeasible) are to be considered. These properties are critical for planning workflows and minimising penalties associated with data movement because scientific data products are asymmetrically distributed across SRCs and replication strategies optimised for performance and cost should be in place when the SRCNet is fully operational.

The costs and impacts associated with data movement and workflow execution encompass the various resources and expenses incurred during the execution of data-intensive operations. 
These primarily include the data transfer (involving energy consumption, bandwidth, and latency when moving data) and the workflow execution  linked to the consumption of computing resources (CPU/GPU hours) and energy. Other key elements are the recurring storage (per TB/month) and the overall processing , which include dedicated hardware, virtualised environments, alongside computing resources used on external sources. It is widely know that computing facilities have increasing needs in terms of power consumption and the cost and impact of data transfer is also significant. The data model has been designed to cover aspects that would allow to take into account green principles.

\section{Advance data analysis services in Science platforms}\label{sec:services}

\subsection{Software as a service  in distributed computing infrastructures}

Science Gateways (SGs) emerged to provide user-friendly environments that allow scientists to build workflows, reuse shared components, and access advanced visualisation tools without needing deep technical expertise \cite{kacsuk2012wspgrade}. To further simplify access and improve efficiency, the Software-as-a-Service (SaaS) paradigm has been adopted in combination with workflow technologies. SaaS enables scientists to focus on the scientific experiment rather than on software installation, configuration, or infrastructure management. % \cite{amaral2014supporting}. 

In this context, we produced a two-level workflow system \cite{Sanchez2016Web} that integrates SaaS principles with the COMPSs programming model \cite{lordan2014servicess}. This system addresses two key challenges: (i) insulating scientists from the complexity of DCIs, and (ii) optimising resource utilisation through parallelism and orchestration.

\begin{figure*}[htbp]
    \centering
    \includegraphics[width=0.75\textwidth]{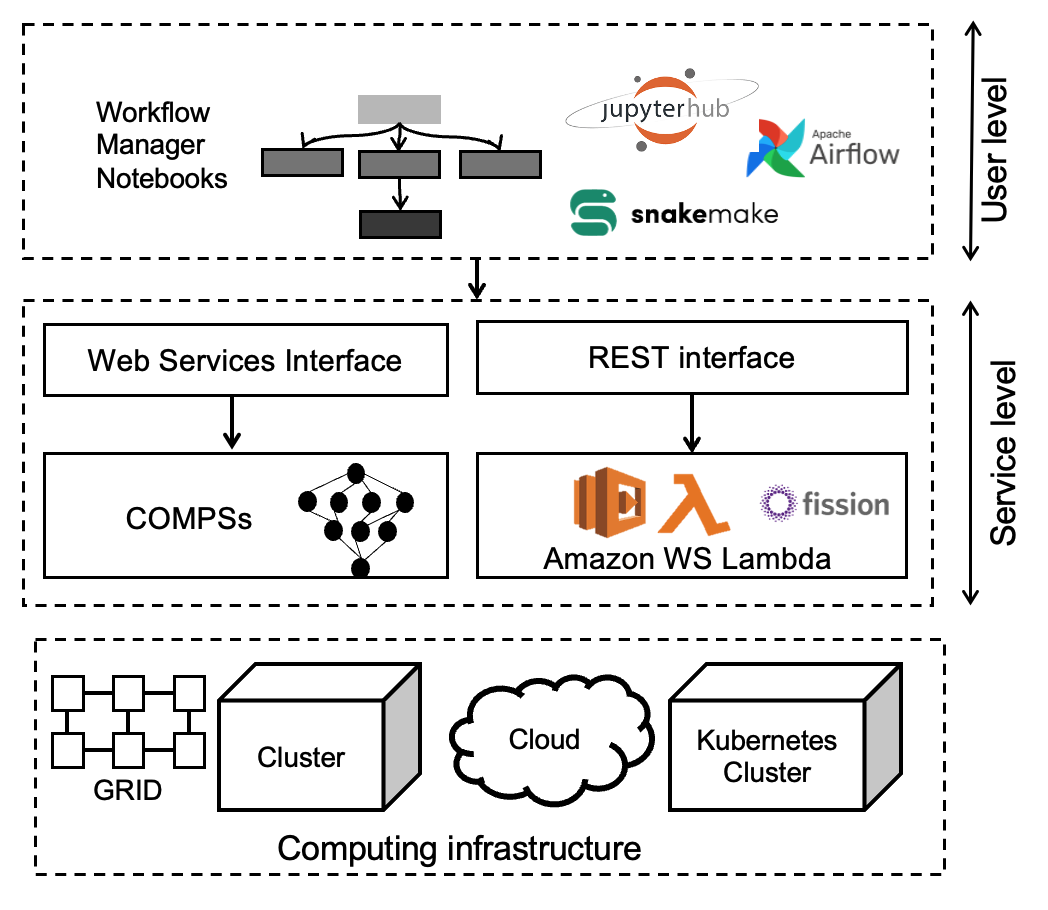}
    \caption{Architecture illustrating how alternative solutions can be combined by advanced analysis tools to enable seamless access to a variety of computing infrastructures.}
    \label{fig:webServices}
\end{figure*}

As seen in Figure \ref{fig:webServices}, the User-Level Workflows are built using web services exposed by scientific applications. These workflows are composed primarily through graphical managers or workflow manager systems, allowing scientists to combine services easily without requiring extensive programming skills. This approach offers significant flexibility and usability, enabling scientists to rapidly create new experiments, integrate domain-specific services with standard Virtual Observatory (VO) protocols, and readily share their processes through platforms or repositories. By abstracting away complex infrastructure details and technical configuration, User-Level Workflows effectively allow scientists to maintain focus purely on achieving their core scientific objectives.

Infrastructure-Level Workflows operate entirely hidden from the end-user, serving to orchestrate the internal execution of tasks within each individual web service using tools like COMPSs. Their primary function is to manage dependencies dynamically while exploiting task-level parallelism inherent in the scientific applications. This allows the system to efficiently submit tasks to the most suitable computing platform available.

The core purpose of these workflows is ensuring highly efficient resource utilisation. They achieve this by maximising the inherent parallelism within applications and significantly minimising data transfer volumes by executing computational tasks as close as possible to the data sources. Furthermore, they are capable of dynamically adapting resource allocation, leveraging elasticity features available in cloud environments to scale resources as necessary.

Because COMPSs automatically detects task dependencies and schedules them to maximise concurrency, these infrastructure workflows reduce overall execution time significantly when compared to traditional sequential processing approaches.

The granularity of web services (defined by the number of tasks grouped within a service) affects both flexibility and performance. Low-granularity services allow fine-grained control and intermediate data manipulation, while high-granularity services reduce overhead and improve execution time by minimising inter-service communication. Performance evaluations confirm that workflows built upon high-granularity services outperform those using atomic services when exploring large parameter spaces. For example, in the kinematical modelling of galaxies, grouping tasks such as ROTCUR and ELLINT into a single service reduced execution time by half compared to workflows invoking these tasks separately \cite{Sanchez2016Web}.

\subsection{Function as a service in the science platform}

Serverless computing abstracts away infrastructure management, allowing developers and scientists to focus on writing and deploying code as discrete functions. These functions are executed in stateless containers, triggered by events such as HTTP requests or workflow invocations. The underlying platform automatically handles resource allocation, scaling, and fault tolerance, offering a pay-as-you-go model that is highly attractive for large-scale projects \cite{andi2021analysis,barrak2022serverless}.

FaaS platforms, such as AWS Lambda, Google Cloud Functions, and open-source solutions like OpenFaaS and Fission, provide lightweight execution environments that can integrate with container orchestration systems (e.g., Kubernetes). This integration enables dynamic scaling and portability across heterogeneous infrastructures, which is essential for federated environments like SRCNet~\cite{malawski2020serverless}.

Radio interferometry workflows typically involve multiple stages: data ingestion, flagging, calibration, imaging, and self-calibration \cite{kale2021capture}. Each stage can be decomposed into modular tasks, which are ideal candidates for implementation as serverless functions. For example, an imaging task could consists of a deconvolution and image reconstruction using \textit{tclean} (CASA) or \textit{WSClean} \cite{vanderTol2018image}. 

We deployed a testbed using Fission \cite{riosmonje2023serverless} on the Spanish Prototype SKA Regional Centre (SPSRC) \cite{garrido2022toward}. The platform runs on a two-node Kubernetes cluster and hosts functions for imaging and calibration. The functions are published via HTTP endpoints and can be invoked from command-line tools or integrated into workflow engines. Preliminary tests indicate minimal overhead compared to native execution, validating the feasibility of this approach.

The serverless model (see Figure \ref{fig:webServices}) offers several key advantages for task execution. One of the most prominent is scalability, facilitated by automatic resource allocation and the system's inherent elasticity, which efficiently handles variable workloads. Furthermore, it provides excellent portability, as functions can be executed across diverse hardware infrastructures without requiring modification. Finally, it enhances interoperability through straightforward integration with APIs, notebooks, and workflow managers.

Despite these benefits, the serverless model still presents significant challenges. These include cold-start latency, which can slow down the initiation of new functions, strict memory limitations imposed on individual functions, and the ongoing need to develop advanced scheduling mechanisms to optimise execution within federated and distributed computing environments.

The initial exploration of serverless computing for radio astronomy workflows has evolved into a fully integrated model within the SKA Regional Centre Network (SRCNet). Unlike the early prototypes that focused on deploying isolated functions on generic FaaS platforms, the current implementation embeds serverless principles into the federated SRCNet architecture, ensuring interoperability, security, and data-proximate execution. This integration leverages both global and site-local services: global services such as SKAO Identity and Access Management (IAM), the SRCNet Permissions API, and Site-Capabilities provide authentication, authorisation, and function discovery across the federation, while local services host containerised functions within Kubernetes clusters co-located with data replicas. This design guarantees that functions are executed as close as possible to the datasets they operate on, reducing network transfers and latency while maintaining compliance with governance policies.

A representative example of this evolution is the deployment of the \textit{gaussconv} function, which performs Gaussian convolution on FITS images. The function is developed using FastAPI to expose an OpenAPI-compliant interface, containerised with all dependencies, and deployed through GitOps workflows powered by Flux CI/CD. Once deployed, the function is registered in the SRCNet Site-Capabilities catalogue and exposed via the GateKeeper service, which enforces federated authentication and authorisation before routing requests to the local Kubernetes cluster. Integration with IVOA DataLink allows users to discover not only where dataset replicas are stored but also which functions are available to process them, enabling seamless coupling of data and computation. This approach transforms serverless computing from a conceptual model into an operational capability within SRCNet, paving the way for federated, reproducible, and FAIR-compliant scientific workflows at SKA scale.

\section{Reproducibility in AMIGA science studies}\label{sec:reprod}
\subsection{Reproducibility in the HCG16 study}

The scientific problem addressed in the HCG16 study \cite{Jones2019_HCG16} concerns the analysis of multi-wavelength data to understand the physical processes in compact galaxy groups. These systems exhibit complex interactions, requiring a combination of observational datasets and computationally intensive modelling to derive robust conclusions. Ensuring reproducibility in such a context is essential, given the heterogeneity of data sources and the complexity of the analysis pipeline.

Reproducibility was approached at two complementary levels: the computational pipeline and the generation of figures and visualisations. Each level required specific strategies to guarantee that other researchers can replicate the results and interact with the analysis without technical barriers.

The first level of reproducibility focused on the main pipeline, which involves computationally intensive tasks such as spectral fitting and image processing. This pipeline was fully containerised, encapsulating all dependencies and configurations within Docker images. By automating the installation of libraries and tools, the pipeline can be deployed on any environment supporting container orchestration, including HPC clusters and cloud platforms. This approach eliminates issues related to software versioning and system compatibility, ensuring that the same workflow can be executed consistently across different infrastructures.

A critical aspect of reproducibility is data availability. The original observational data were initially stored in a restricted archive requiring manual access. To overcome this limitation, the datasets were published in the EUDAT repository, providing persistent identifiers and open access. Furthermore, the derived data products (e.g. processed images and spectral cubes) were deposited in cloud storage and registered in Virtual Observatory (VO) repositories, enabling discovery through standard protocols and integration with VO-compliant tools.

The second level of reproducibility addressed the generation of figures and plots presented in the study. These visualisations are not static outputs but integral components of the scientific interpretation. To ensure transparency and interactivity, all figure-generation scripts were implemented in Jupyter Notebooks. This choice allows researchers to inspect, modify, and re-execute the code that produced each figure.

To facilitate access without requiring local installation of dependencies, the notebooks were integrated with Binder. This service dynamically provisions an execution environment in the cloud, enabling users to run the notebooks directly from a web browser. Combined with Conda environments, which specify all required packages and versions, this setup guarantees that the computational context is fully reproducible. Users can not only reproduce the figures but also experiment with alternative parameters or datasets, promoting an open and collaborative approach to scientific analysis. In fact, it can be launched directly in myBinder and in the European Grind Infrastructure (EGI). 

The importance of these tools lies in their ability to bridge reproducibility and usability. By leveraging Jupyter, Binder, and Conda, the study provides an interactive platform where figures can be regenerated and customised without technical overhead. This design aligns with the principles of Open Science, ensuring that results are transparent, verifiable, and extensible.

\begin{figure*}[tbp]
    \centering
    \includegraphics[width=0.85\textwidth]{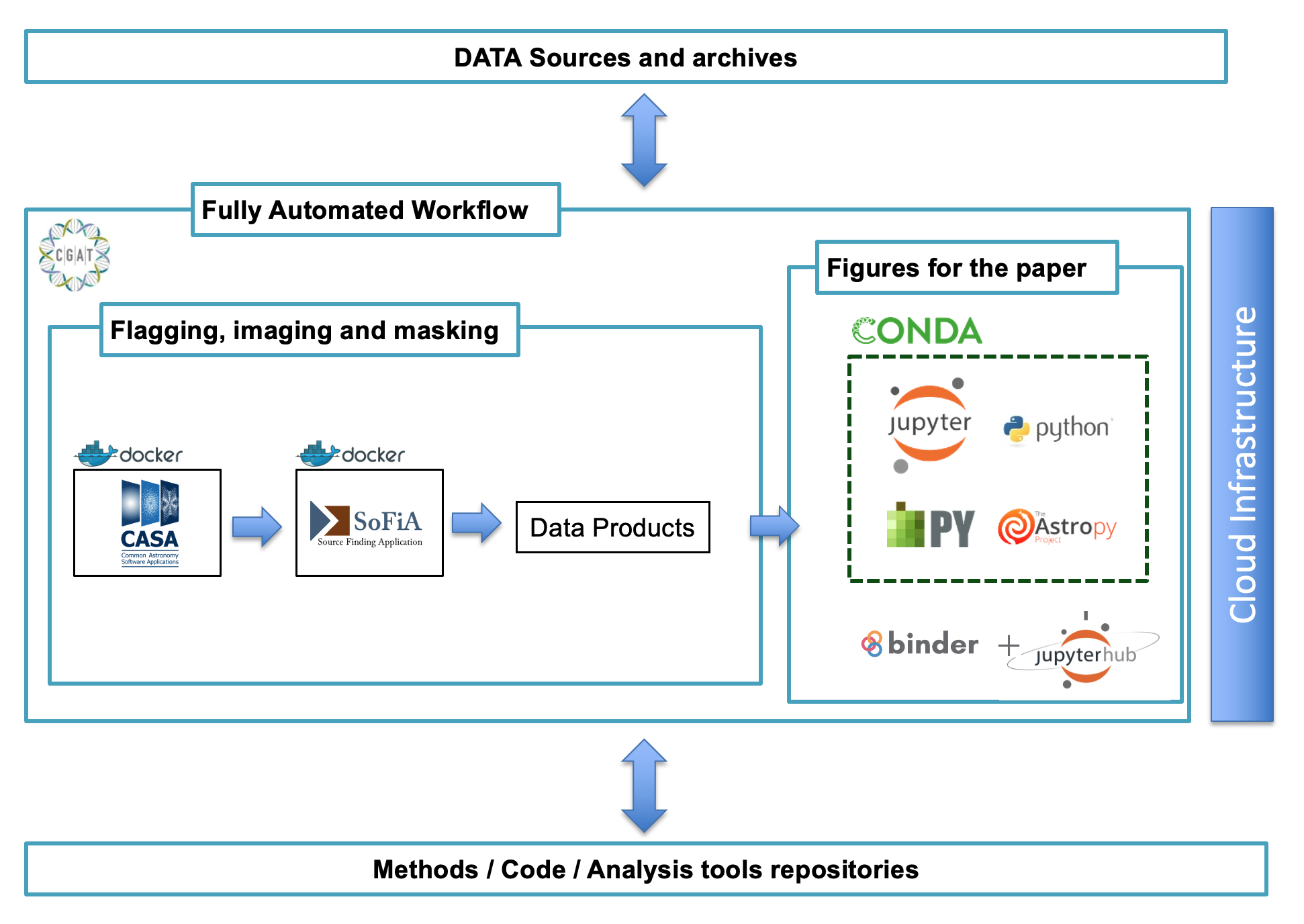}
    \caption{Architecture for reproducibility in the HCG16 study. The top layer represents data sources and scientific archives for original and derived datasets. The bottom layer includes repositories for methods and code (e.g., GitHub, container registries). The middle layer is divided into two components: reproducibility of the computational pipeline and reproducibility of figures. The bottom layer shows the underlying infrastructures (cloud, grid, HPC, Kubernetes) supporting deployment and execution.}
    \label{fig:reproducibility-HCG16}
\end{figure*}

A comprehensive description of the methods, tools, and instruments used to achieve reproducibility is available in the public GitHub repository \cite{mike_jones_2024_10977609}, which includes container definitions, workflow scripts, and Jupyter notebooks. Figure~\ref{fig:reproducibility-HCG16} summarises the architecture adopted to guarantee reproducibility at both levels.

\subsection{Reproducibility in a MeerKAT study on HCGs}
%\subsection{Scientific Context}
The work presented at \cite{Ianja2025a} addresses the analysis of Hi gas in six Hickson Compact Groups (HCGs) observed with the MeerKAT radio telescope. These systems are key laboratories for studying galaxy interactions and the role of environment in gas depletion. The observations, part of an approved MeerKAT proposal, produced approximately 50~TB of raw data, making this dataset representative of the challenges posed by next-generation facilities such as the SKA.

The reduction pipeline transforms raw interferometric data into calibrated and imaged products suitable for scientific interpretation. This involves iterative flagging, cross-calibration, self-calibration, and spectral line imaging, complemented by source finding with \texttt{SoFiA}. Each stage requires visual inspection and parameter refinement, ensuring that the final calibration reflects the best achievable quality, while preserving the final steps and parameters.

Again, ensuring reproducibility in this context is non-trivial due to the heterogeneous nature of radio data processing tools and the large volume of intermediate products. The approach in this work consists of integrating workflow management, containerisation, and version control to guarantee that every step (from raw data ingestion to final plots) can be replicated on different infrastructures. 

A central component of this reproducibility strategy in this research is the use of \texttt{Snakemake} \cite{snakemake2012}, a workflow engine that enables the formalisation of the entire data reduction and analysis pipeline. The workflow is expressed as a set of rules, each defining input, output, and computational steps. This declarative structure allows automatic resolution of dependencies and parallel execution. Notably, our pipeline includes not only calibration and imaging scripts but also the generation of diagnostic plots and even the compilation of the manuscript in PDF format, ensuring end-to-end reproducibility.

The adoption of \texttt{Snakemake}  provides additional benefits such as portability and scalability. By abstracting execution details, the same workflow can run on local machines, HPC clusters, or cloud platforms without modification. This flexibility is essential for projects dealing with data volumes of the order of tens of terabytes, as in the present case.

Complementing the workflow engine, we employ \texttt{Stimela}, a container-based framework tailored for radio astronomy. \texttt{Stimela} encapsulates widely used packages such as CASA, WSClean, Cubical, and SoFiA within Docker or Singularity containers, ensuring consistent environments across executions. This approach eliminates dependency conflicts and facilitates reproducibility by fixing software versions and configurations.

On top of \texttt{Stimela}, the \texttt{CARACAL} pipeline orchestrates the calibration and imaging steps specific to MeerKAT data. \texttt{CARACAL} \cite{Jozsa2020} acts as a Python wrapper that chains multiple applications into a coherent sequence, passing intermediate products between stages. This modular design allows transparent integration with \texttt{Snakemake}, so that each CARACAL task becomes a reproducible rule within the global workflow.

In addition, an adapted version of the MeerKAT pipeline\footnote{processMeerKAT: \url{https://github.com/manuparra/meerkat-pipeline-custom-hpc}} has been developed to enable its run on arbitrary high-performance computing (HPC) environments and clusters, without relying on the original pipeline’s site-specific dependencies. This adaptation standardises not only the pipeline configuration but also the underlying software stack, allowing for deployment on custom computational resources. As a result, the reproducibility of the data-processing workflow is significantly improved, while maintaining flexibility to operate across heterogeneous HPC infrastructures. This work also required the installation, deployment, and configuration of a workload management system based on Slurm, which is a mandatory component for pipeline execution.

Version control plays a critical role in tracking changes to scripts, configuration files, and parameter sets. All workflow definitions and container recipes are maintained in a repository, enabling external verification and reuse. Parameter files corresponding to the final calibration are preserved alongside intermediate logs, providing a complete provenance record.

Finally, reproducibility extends to data management. Given the initial 50~TB of raw visibilities, we implement strategies to minimise data transfer by selecting sub-datasets at the origin. Each target is processed individually, reducing the working set to approximately 100~GB per run, with intermediate products reaching 500~GB before compression to a final 10~GB science-ready dataset. %These figures illustrate the computational and storage requirements that future SKA-scale projects must anticipate.

\section{Conclusions}\label{sec:conclusions}

The SKAO will face technological challenges in order to ensure that the science it enables is  scalable and reproducible across heterogeneous computing environments and over long timescales. Reproducibility at the scale of the SKA cannot be treated as an afterthought, but must instead be addressed as a first-order requirement, deeply embedded in data models, software architectures and operational practices. In this context, the work reviewed in this paper highlights the necessity of combining advances in data modelling, workflow preservation and distributed computing in order to support reproducible and verifiable scientific outcomes.

This paper has presented a review of a selection of research activities undertaken by the AMIGA group that address key aspects of these challenges. From the perspective of data and knowledge representation, semantic and model-based approaches provide the foundations for describing data products, workflows and their dependencies in a machine-actionable manner. Complementarily, the development of data analysis services deployed over distributed and heterogeneous computing infrastructures demonstrates how complex scientific workflows can be orchestrated and executed in a reproducible way, despite the intrinsic diversity of execution environments. Together, these approaches illustrate how reproducibility can be improved through the integration of data models, software services and computing infrastructures.

Finally, the application of current technologies to two astronomy use cases based on SKA precursor and pathfinder data has demonstrated that viable ecosystems for reproducible science already exist today. These results provide concrete evidence that reproducibility is achievable in practice when appropriate tools and methodologies are adopted. However, the SKA scale requires that such capabilities are designed into the system from the beginning, particularly within the architecture of the SRC Network. In this regard, the espSRC is incorporating Open Science principles transversally across its design and development activities. This design-driven approach will be essential for maximising the scientific return of the SKAO in the long term.

\section*{Acknowledgements}

The authors acknowledge financial support from the grant CEX2021-001131-S funded by MICIU/AEI/ 10.13039/501100011033 and from the grant TED2021-130231B-I00 funded by MICIU/AEI/ 10.13039/501100011033 and by the European Union NextGenerationEU/PRTR, acknowledges financial support from the grant  PID2021-123930OB-C21, PID2024-155817OB-I00 funded by MICIU/AEI/ 10.13039/501100011033 and by ERDF/EU, and by the grant INFRA24023 (CSIC4SKA) funded by CSIC. The authors acknowledges the Prototype of an SRC (SPSRC) service and support funded by the Ministerio de Ciencia, Innovación y Universidades (MICIU), by the Junta de Andalucía, by the European Regional Development Fund (ERDF) and by the European Union NextGenerationEU/PRTR. The SPSRC acknowledges financial support from the Agencia Estatal de Investigación (AEI) through the "Center of Excellence Severo Ochoa" award to the Instituto de Astrofísica de Andalucía (IAA-CSIC) (SEV-2017-0709) and from the grant CEX2021-001131-S funded by MICIU/AEI/ 10.13039/501100011033.
%
% ---- Bibliography ----
%

\bibliographystyle{spbasic}
\bibliography{References}

%\begin{thebibliography}{6}
%

%\bibitem {smit:wat}
%Smith, T.F., Waterman, M.S.: Identification of common molecular subsequences.
%J. Mol. Biol. 147, 195?197 (1981). \url{doi:10.1016/0022-2836(81)90087-5}

%\bibitem {may:ehr:stein}
%May, P., Ehrlich, H.-C., Steinke, T.: ZIB structure prediction pipeline:
%composing a complex biological workflow through web services.
%In: Nagel, W.E., Walter, W.V., Lehner, W. (eds.) Euro-Par 2006.
%LNCS, vol. 4128, pp. 1148?1158. Springer, Heidelberg (2006).
%\url{doi:10.1007/11823285_121}

%\bibitem {fost:kes}
%Foster, I., Kesselman, C.: The Grid: Blueprint for a New Computing Infrastructure.
%Morgan Kaufmann, San Francisco (1999)

%\bibitem {czaj:fitz}
%Czajkowski, K., Fitzgerald, S., Foster, I., Kesselman, C.: Grid information services
%for distributed resource sharing. In: 10th IEEE International Symposium
%on High Performance Distributed Computing, pp. 181?184. IEEE Press, New York (2001).
%\url{doi: 10.1109/HPDC.2001.945188}

%\bibitem {fo:kes:nic:tue}
%Foster, I., Kesselman, C., Nick, J., Tuecke, S.: The physiology of the grid: an open grid services architecture for distributed systems integration. Technical report, Global Grid
%Forum (2002)

%\bibitem {onlyurl}
%National Center for Biotechnology Information. \url{http://www.ncbi.nlm.nih.gov}

%\end{thebibliography}
\end{document}